\documentstyle[aps,prb,psfig,multicol]{revtex}
\begin{document}

\newcommand{\uprule}{\end{multicols}}
\newcommand{\downrule}
{\begin{multicols}{2}\narrowtext}
\draft
\sloppy

\title{Triple approach to determination of the c-axis penetration depth
in BSCCO crystals}

\author{M.~R.~Trunin, Yu.~A.~Nefyodov, D.~V.~Shovkun, and A.~A.~Zhukov,}
\address{Institute of Solid State Physics, 142432, Chernogolovka,
Moscow district, Russia}

\author{N.~Bontemps}
\address{Laboratoire de Physique du Solide ESPCI, 10 rue Vauquelin,
75231 Paris cedex 05, France}

\author{A.~Buzdin and M.~Daumens}
\address{Laboratoire de Physique Th\'eorique, Universit\'e Bordeaux I,
33405 Talence Cedex, France}

\author{H.~Enriquez}
\address{SRSIM/DRECAM/DSM, CEA-Saclay, 91191 Gif sur Yvette cedex, France}

\author{T.~Tamegai}
\address{Department of Applied Physics, The University of Tokyo,
Hongo, Bunkyo-ku, 113-8656, Japan}

\maketitle

\begin{abstract}
The $c$-axis penetration depth $\lambda_c$ in
Bi$_2$Sr$_2$CaCu$_2$O$_{8+\delta}$ (BSCCO) single crystals as a
function of temperature has been determined using three high-frequency
techniques, namely: (i) measurements of the ac-susceptibility at a
frequency of 100~kHz for different sample alignments with respect to
the ac magnetic field; (ii) measurements of the surface impedance in
both superconducting and normal states of BSCCO
crystals at 9.4~GHz; (iii) measurements of the surface barrier field
$H_J(T)\propto 1/\lambda_c(T)$ at which Josephson vortices penetrate
into the sample. Careful analysis of these measurements, including
both numerical solution of the electrodynamic problem of the
magnetic field distribution in an anisotropic plate at an arbitrary
temperature and influence of defects in the sample, has allowed us
to estimate $\lambda_c(0)\approx 50$~$\mu$m in BSCCO crystals overdoped
with oxygen ($T_c\approx 84$~K) and $\lambda_c(0)\approx 150$~$\mu$m at
the optimal doping level ($T_c\approx 90$~K). The results obtained by
different techniques are in reasonable agreement.

\end{abstract}

\vspace{0.4cm}


\vspace{0.4cm}

\begin{multicols}{2}

\section{Introduction}

Study of anisotropy of the
magnetic field penetration depth as a function of temperature
in high-$T_c$ superconductors (HTS) advance considerably our
understanding of pairing mechanism in these materials.
It is known (see, e.g., Ref.~\cite{Tru1} and references therein) that
in-plane penetration depth $\Delta\lambda_{ab}(T)\propto T$ in the range
$T<T_c/3$ in high-quality HTS samples at the optimal level of doping,
and this observation can be interpreted the most simply in the
$d$-wave model of the high-frequency response of HTS.
Measurements of out-of-plane or $c$-axis penetration depth
$\lambda_c(T)$ are quoted less frequently than
those of $\lambda_{ab}(T)$. Most of such data published so far
were derived from microwave measurements of the surface impedance of HTS
crystals \cite{Shib1,Mao,Kit1,Bon1,Jac1,Shib2,Srik,Kit2,Hos}. There is
no consensus about $\Delta\lambda_c(T)$ at low temperatures. Even in
high-quality YBCO crystals, which are the most studied objects, one can
find both linear, $\Delta\lambda_c(T)\propto T$ \cite{Mao,Srik}, and
quadratic dependences \cite{Hos} in the range $T<T_c/3$. In BSCCO
materials, the shape of $\Delta\lambda_c(T)$ depends on the level of
oxygen doping: in samples with maximal $T_c\simeq 90$~K
$\Delta\lambda_c(T)\propto T$ at low temperatures \cite{Jac1,Shib2}; at
higher oxygen contents (overdoped samples) $T_c$ is lower and the
linear function $\Delta\lambda_c(T)$ transforms to a quadratic one
\cite{Shib2}. The common feature of all microwave experiments is that
the change in the ratio $\Delta\lambda_c(T)/\lambda_c(0)$ is smaller
than in $\Delta\lambda_{ab}(T)/\lambda_{ab}(0)$ because in all HTS
$\lambda_c(0)\gg\lambda_{ab}(0)$.
Another possibility to determine $c$-axis penetration depth is
the measurements of the first penetration field of Josephson
vortices. In quasi-2D systems, their penetration may be impeded by a surface
barrier, the value of which is inversely proportional to
$\lambda_{c}(T)$ \cite{buzdin1}. The quantitative estimates of
$\lambda_{c}(T)$ deduced from the surface barrier data were however
disputed \cite{hussey}.
Furthermore, $\lambda_{c}(T)$ is also inversely proportional to the
plasma frequency \cite{bulaevskii,koshelev} which is
usually assigned to Josephson plasma resonance frequency
modified by the field-dependent interlayer phase coherence
\cite{bulaevskii,matsuda2}. However, this interpretation is still
controversial \cite{sonin}.
Therefore, independent measurements of $\lambda_c(T)$
are of interest. To date, all the above mentioned properties have been
studied separately. It is the aim of this paper to apply together
three different techniques to the determination of the absolute value
of $\lambda_c(T)$ in order to obtain unambiguous results:
(i) ac-susceptibility measurements of BSCCO crystals have allowed us
to determine the temperature variation $\Delta \lambda_c(T)$;
ii) cavity perturbation technique and electrodynamic analysis of the
surface impedance anisotropy is used to determine both the variations
and absolute values of $\lambda_c(T)$ and $\lambda_{ab}(T)$;
(iii) the first penetration field of Josephson
vortices is measured and shown to be related to $\lambda_c(T)$.

\section{Electrodynamic basis of the measurements}

The electrodynamics of layered anisotropic HTS
is characterized by the components $\sigma_{ab}$ and $\sigma_c$
of the conductivity tensor. In the normal state, ac field penetrates
in the direction of the $c$-axis through the skin depth
$\delta_{ab}=\sqrt{2/\omega\mu_0\sigma_{ab}}$ and in the CuO$_2$ plane
through $\delta_c=\sqrt{2/\omega\mu_0\sigma_c}$. In the
superconducting state all parameters $\delta_{ab}$, $\delta_c$,
$\sigma_{ab}=\sigma'_{ab}-i\sigma''_{ab}$, and $\sigma_c=\sigma'_c
-i\sigma''_c$ are complex. In the temperature range $T<T_c$, if
$\sigma'\ll \sigma''$, the field penetration depths are given by the
formulas $\lambda_{ab}=\sqrt{1/\omega\mu_0\sigma''_{ab}}$,
$\lambda_c=\sqrt{1/\omega\mu_0\sigma''_c}$. In the close neighborhood of
$T_c$, if $\sigma'> \sigma''$, the decay of magnetic field in
the superconductor is characterized by the functions
$\rm {Re}\,(\delta_{ab})$ and $\rm {Re}\,(\delta_c)$, which turn to
$\delta_{ab}$ and $\delta_c$ at $T\ge T_c$, respectively.

In this paper we present the results of high-frequency measurements
of anisotropy of two BSCCO single crystals with various levels of oxygen
doping. The first sample (\#1), characterized by a lower critical
temperature, $T_c\approx 84$~K (slightly overdoped), has dimensions
$a\times b\times c\simeq 0.8\times 1.8\times 0.03$~mm$^3$.
The second (\#2, $a\times b\times c\simeq 1.5\times 1.5\times 0.1$~mm$^3$)
is almost optimally doped ($T_c\approx 90$~K).

\begin{figure}[h]
\centerline{\psfig{figure=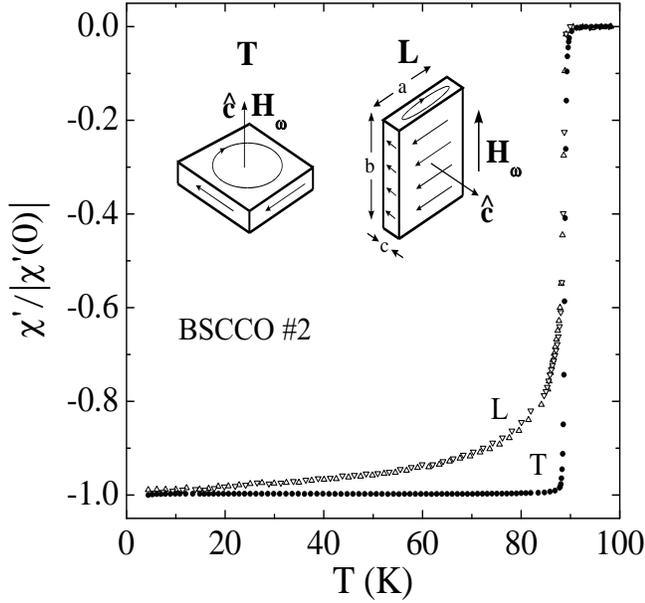,height=8cm,width=8.5cm,clip=,angle=0.}}
\downrule
\caption{Curves of the ac-susceptibility of sample~\#2 versus
temperature in different orientation with respect to ac magnetic field:
${\bf H}_{\omega}\parallel {\bf c}$ (full circles);
${\bf H}_{\omega}\perp {\bf c}$, ${\bf H}_{\omega}$ is parallel to
the $b$-edge of the crystal (up triangles);
${\bf H}_{\omega}\perp {\bf c}$, ${\bf H}_{\omega}$ is
parallel to the $a$-edge of the crystal (down triangles).
Left-hand inset: transverse (T) orientation,
${\bf H}_{\omega}\parallel {\bf c}$, the arrows on the surfaces show
directions of the screening current. Right-hand inset: Longitudinal (L)
orientation, ${\bf H}_{\omega}\perp {\bf c}$.}
\uprule
\label{f1}
\end{figure}

For determination of both $\lambda_{ab}(T)$ and $\lambda_c(T)$ components
of the penetration depth we measured the temperature dependences of the
Q-factor and the frequency shift $\delta f$ of the resonant circuit
for different sample alignments with respect to the ac magnetic field
${\bf H}_{\omega}$: in the transverse (T) orientation,
${\bf H}_{\omega}\parallel {\bf c}$, when the screening current flows
in the $ab$-plane of the crystal and in the
longitudinal (L) orientation, ${\bf H}_{\omega}\perp {\bf c}$,
with currents running in the directions of both CuO$_2$ planes
and the $c$-axis (insets to Fig.~1). In the first case the values of Q(T)
and $\delta f(T)$ are directly connected with in-plane penetration
depth $\lambda_{ab}(T)$ at $T<T_c$ and skin depth $\delta_{ab}(T)$ at
$T\ge T_c$ \cite{Tru1}. Both lengths are smaller than the thickness of
the crystal. In the second L-orientation at $T<0.9\,T_c$ the penetration
depth is still smaller than characteristic sample dimensions.
But at $T>0.9\,T_{\rm c}$ the lengths $\lambda_c$ and $\delta_c$ are
comparable to the width of the crystal. In order to analyze our
measurements in both superconducting and normal states, we used
formulae for field distributions in an anisotropic long strip
($b\gg{a,c}$) in the L-orientation. These formulae neglect the effect of
$ac$-faces of the crystal, if ${\bf H}_{\omega}$ is parallel to the
$b$-edge of the crystal (inset on the right of Fig.~1), but take into account
 the size effect. At an arbitrary temperature,
the measured quantities are expressed in terms of the complex parameter
$\mu$ introduced in Ref.~\cite{Gou}, which is controlled by the components
$\sigma_{ab}(T)$ and $\sigma_c(T)$ of the conductivity tensor
through the penetration depths $\delta_{ab}$ and $\delta_c$ \cite{Tru90}:
{\scriptsize
\begin{eqnarray}
\mu={8 \over \pi^2}\sum_n {1\over n^2}\left\{
{\tan(\alpha_n) \over \alpha_n}+
{\tan(\beta_n) \over \beta_n}
\right\}, \nonumber \\
\alpha_n^2=-{a^2 \over \delta_c^2}
\left({i\over 2}+{\pi^2 \over 4}{\delta_{ab}^2 \over c^2}n^2 \right),\,
\beta_n^2=-{c^2 \over \delta_{ab}^2}
\left({i\over 2}+{\pi^2 \over 4}{\delta_c^2 \over a^2}n^2 \right),
\end{eqnarray}}
where the sum is performed over odd integers $n>0$. In the
superconducting state, if $\sigma'\ll \sigma''$, the parameter $\mu$ is real:
{\scriptsize
\begin{eqnarray}
\mu={8 \over \pi^2}\sum_n {1\over n^2}\left\{
{\tanh(\tilde\alpha_n) \over \tilde\alpha_n}+
{\tanh(\tilde\beta_n) \over \tilde\beta_n}
\right\}, \nonumber \\
\label{eq3}
\tilde\alpha_n^2={a^2 \over \lambda_c^2}
\left({1 \over 4}+{\pi^2 \over 4}{\lambda_{ab}^2 \over c^2}n^2 \right),
\quad
\tilde\beta_n^2={c^2 \over \lambda_{ab}^2}
\left({1 \over 4}+{\pi^2 \over 4}{\lambda_c^2 \over a^2}n^2 \right).
\end{eqnarray}}

\section{ac-susceptibility measurements}

The first approach to determination of the $c$-axis penetration depth
$\lambda_c(T)$ in BSCCO single crystals is based on measuring the
ac-susceptibility $\chi=\chi'-i\chi''$ at a frequency of $100$~kHz.
The imaginary part
$\chi''$ of $\chi$ is proportional to the energy dissipation in the sample
and the real part $\chi'$ is proportional to the shielding of magnetic field.
Ac-susceptibility is characterized by the components of the conductivity
tensor and, hence, allows to determine the penetration depth $\lambda$.
Experimentally it is possible if the penetration depth is comparable to
the sample dimension. So the ac-susceptibility measurements are
usually adapted to the superconducting powders whose grain size is
comparable to $\lambda$. However, since $\lambda_c(0)$ in BSCCO single
crystals is relatively large, we managed to determine $\lambda_c(T)$
from the temperature dependences of $\chi'(T)$.

As an example, figure~1 shows the temperature dependences
$\chi'(T)/|\chi'(0)|$ in
sample~\#2 for three different sample alignments with respect to
the ac magnetic field: in the T-orientation,
${\bf H}_{\omega}\parallel {\bf c}$, (full circles); in the
L-orientation, ${\bf H}_{\omega}\perp {\bf c}$,
(${\bf H}_{\omega}$ is parallel to the $b$-edge of the crystal,
up triangles); in the L-orientation,
whose difference from the previous configuration is that the sample is
turned around the $c$-axis through $90^\circ$ (down triangles).
Fig.~1 clearly shows that at
$T<T_c$ $\chi'_{ab}(T)$ is notably smaller in the T-orientation than
$\chi'_{ab+c}(T)$ in the L-orientation (the subscripts of $\chi'$ denote
the direction of the screening current). The coincidence
of $\chi'_{ab+c}(T)$ curves at ${\bf H}_{\omega}\perp {\bf c}$
and the small width of the superconducting transition at
${\bf H}_{\omega}\parallel {\bf c}$
($\Delta T_c< 1$~K) indicate that the quality of the tested
sample~\#2 is fairly high.

In the superconducting state at $T<0.9\,T_c$ we find that
$\lambda_{ab}\ll c$ and $\lambda_c\ll a$. In this case, we derive from
Eq.~(2) a simple relation between the real parts of $\mu$ and $\chi$:
\begin{equation}
\mu'=1+\chi'={2\lambda_c \over a}+ {2\lambda_{ab} \over c}~.
\end{equation}

\begin{figure}[h]
\centerline{\psfig{figure=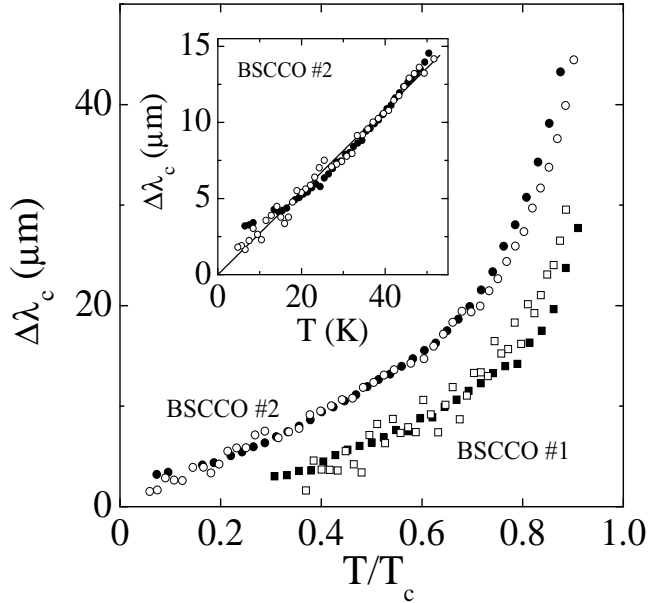,height=8cm,width=8.5cm,clip=,angle=0.}}
\downrule
\caption{Temperature dependences $\Delta\lambda_c$
in sample~\#1 (squares) and \#2 (circles) at $T<0.9\,T_c$. Open
symbols plot low-frequency measurements, full symbols show
microwave data. The inset shows low temperature sections of the
$\Delta\lambda_c$ curves in sample~\#2.}
\uprule
\label{f2}
\end{figure}

Figure~2 shows measurements of $\Delta\lambda_c(T)$ in
sample~\#1 (squares) and sample~\#2 (circles) at $T<0.9\,T_c$. The
open symbols plot low-frequency measurements obtained from $\chi'(T)$
in accordance with Eq.~(3), the full symbols plot microwave measurements
(see the next section). Agreement between these measurements is fairly
good, but in fitting together experimental data from sample~\#2
(upper curve) we had to divide by a factor of 1.8 all
$\Delta\lambda_c(T)$ derived from measurements of ac-susceptibility
using Eq.~(3). The cause of the difference between $\Delta\lambda_c(T)$
measured in sample~\#2 at different frequencies is not quite clear
\cite{Tru90}.

The curves of $\Delta\lambda_c(T)$ at $T<0.5\,T_c$ plotted in Fig.~2
are almost linear: $\Delta\lambda_c(T)\propto T$. The inset to Fig.~2
shows the low-temperature section of the curve of $\Delta\lambda_c(T)$
in sample~\#2. Its slope is 0.3~$\mu$m/K and equals that from
Ref.~\cite{Shib2}. Note that changes in $\Delta\lambda_c(T)$ are
smaller in the oxygen-overdoped sample~\#1 than in sample~\#2.

We also estimated $\lambda_c(0)$ on the base of absolute
measurements of the susceptibility $\chi'_c(0)$ and we
obtained $\lambda_c(0)\approx 70$~$\mu$m for sample~\#1 and
$\lambda_c(0)\approx 210$~$\mu$m for sample~\#2.

\section{cavity perturbation technique}

The second technique of determination $\lambda_c(T)$ is measuring
the difference $\Delta(1/Q)$ ($\Delta f$) between reciprocal Q's
(resonant frequency shifts) of a
cavity with a sample inside and the empty cavity as functions of
temperature at a frequency $f=9.4$~GHz. These parameters are related to
the surface impedance $Z_s=R_s+iX_s$ components through the geometrical
factor $\Gamma_{\rm s}$ of the sample: $R_s=\Gamma_s\Delta(1/Q)$,
$\Delta X_s=-2\Gamma_s\Delta f/f$ \cite{Tru1}.
The penetration depth is $\lambda (T)=X_s(T)/\omega\mu_0$ at $T<T_c$.
In order to determine
the absolute value of $X_s=-2\Gamma_s\delta f/f$, where $\delta f$
is the frequency shift relative to that which would be measured
for a sample with perfect screening and no penetration of the microwave
fields, one needs the constant parameter $f_0=\Delta f(T)-\delta f(T)$.
In HTS this constant can be derived from microwave measurements in the
normal state.

In the T-orientation $f_0$ can be derived from the condition that the
real and imaginary parts of the impedance should be equal above $T_c$
(normal skin-effect) \cite{exp}. Given $f_0$, we can calculate the
conductivity
$\sigma_{ab}(T)=\sigma'_{ab}-i\sigma''_{ab}=i\omega\mu_0/Z_s^2(T)$ at
all $T$. The temperature dependences of $\sigma'_{ab}(T)$ (open squares)
and $\sigma''_{ab}(T)$ (open circles) are shown
in Fig.~3a and Fig.~3b for crystals~\#1 and \#2 respectively. Note
that $\sigma'_{ab}(T)$ in Fig.~3b does not have a broad peak at low
temperatures, typical for $\sigma'_{ab}(T)$ in high-quality HTS. The
reason for that is the rather large value of the residual surface resistance:
$R_{\rm res}\equiv R_s(T\to 0) \approx 0.5$~m$\Omega$ in the $ab$-plane
of the sample \#2 \cite{Tru10}, $R_{\rm res}$ was
four times less in the sample \#1.

\uprule
\widetext
\begin{figure}[t]
\centerline{\psfig{figure=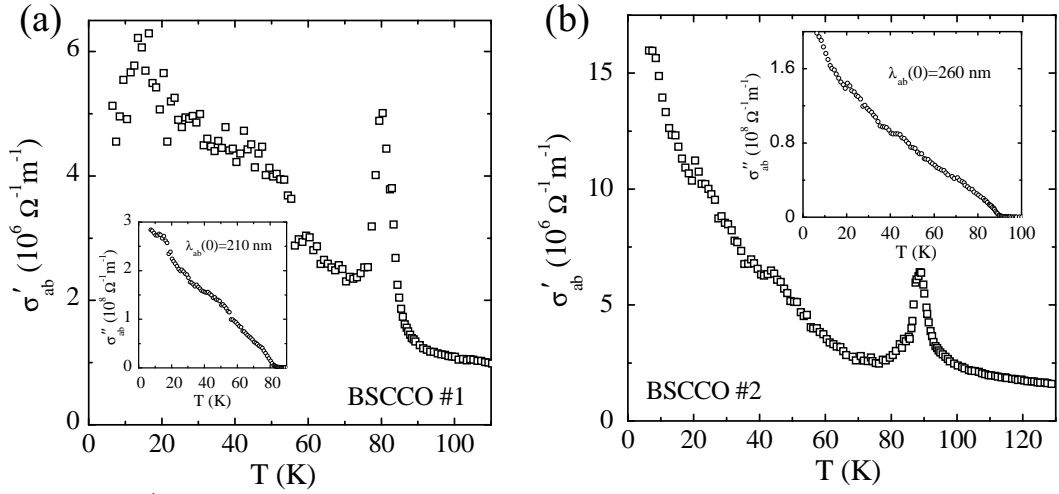,height=6.5cm,width=14cm,clip=,angle=0.}}
\caption{Conductivities $\sigma'_{ab}(T)$ (open squares) of two BSCCO
single crystals \#1 (fig.~a) and \#2 (fig.~b) at 9.4~GHz, extracted
from the surface impedance measurements. The insets show
$\sigma''_{ab}(T)=\lambda_{ab}^2(0)/\lambda_{ab}^2(T)$ data
(open circles).}
\label{3}
\end{figure}
\downrule

In the L-orientation of a crystal shaped as a long strip
the quantities $\Delta(1/Q)$ and $\delta f$ are expressed in terms
of the complex function $\mu=\mu'-i\mu''$ from Eq.~(1) or
$\chi=(-1+\mu)$:

\begin{equation}
\Delta\,(1/Q)-2i\,\delta f/f=i \gamma \mu v/V,
\end{equation}
where $v$ and $V=58$~cm$^3$ are the volumes of the sample and cavity
respectively, $\gamma=10.6$ is a constant of our cavity \cite{Tru1}.
In the superconducting state at $T<0.9\,T_c$ the expression for $\mu$
is given by Eq.~(3). The curves of $\Delta\lambda_c(T)$ in sample~\#1
(full squares) and sample~\#2 (full circles), measured by cavity
perturbation technique and obtained using Eqs.~(3), (4), are shown in Fig.~2.

\uprule
\widetext
\begin{figure}[t]
\centerline{\psfig{figure=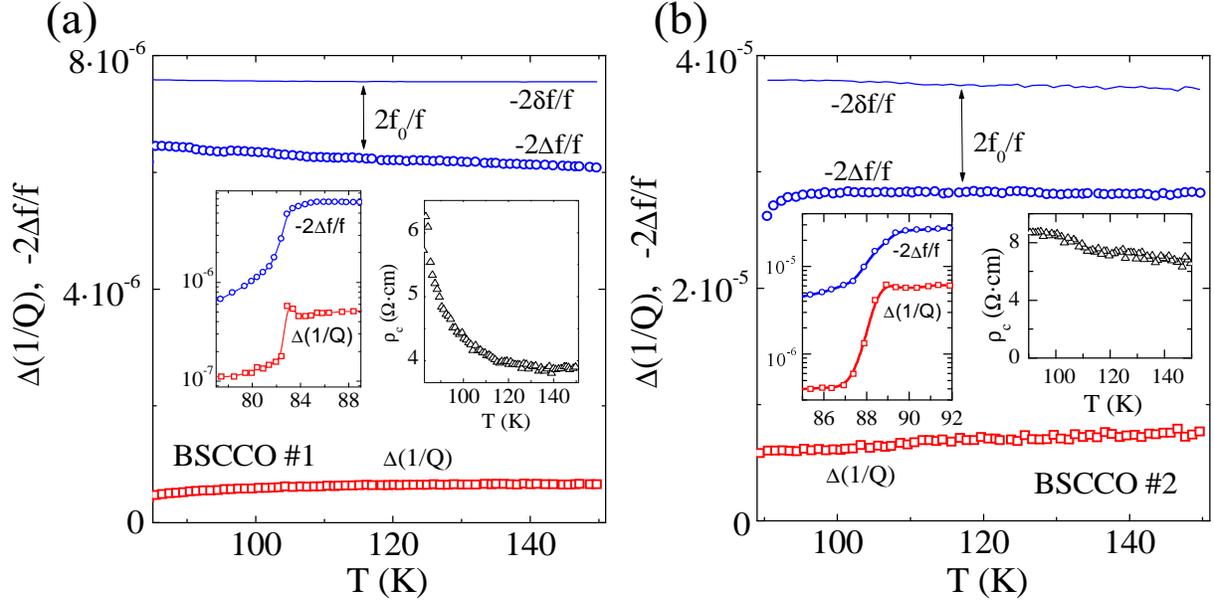,height=8cm,width=16cm,clip=,angle=0.}}
\caption{Temperature dependences of $\Delta (1/Q)$ (open squares) and
$-2\Delta f/f$ (open circles) at ${\bf H}_{\omega}\parallel {\bf b}$
of sample~\#1 (fig.~a) and sample \#2 (fig.~b) at $T>T_c$. Solid lines
show the functions $-2\delta f(T)/f$ deriving from Eqs.~(4) and (1).
Left-hand insets: $\Delta(1/Q)$ and $-2\Delta f/f$ as
functions of temperature in the neighborhood of $T_c$. Right-hand insets:
$\rho_c(T)$ (triangles).}
\label{4}
\end{figure}
\downrule

We can estimate $\lambda_c(0)$ by comparing of $\Delta(1/Q)$ and
$\Delta f=\delta f-f_0$
measurements taken in the T- and L-orientations to numerical calculations
by Eqs.~(1) and (4), which take account of the size effect in
the high-frequency response of an anisotropic crystal. The procedure of
comparison for sample~\#1 and \#2 is illustrated by Fig.~4a and Fig.~4b
respectively. Unlike the case of
the T-orientation, the measured temperature dependence of $\Delta(1/Q)$
in the L-orientation deviates from $(-2\Delta f/f)$ owing to the size
effect. Using the measurements of $\sigma_{ab}(T)$ at $T>T_c$ in the
T-orientation (Fig.~3), alongside
the data on $\Delta(1/Q)$ in the L-orientation (open squares in Fig.~4),
from Eqs.~(1) and (4) we obtain the curves of $\rho_c(T)=1/\sigma_c(T)$
shown in the right-hand insets to Fig.~4a and Fig.~4b for crystals \#1
and \#2. Further, using the functions $\sigma_c(T)$ and $\sigma_{ab}(T)$,
we calculate $(-2\delta f/f)$ versus
temperature for $T>T_c$, which are plotted by the solid lines in Fig.~4.
These lines are approximately parallel to the experimental curves of
$-2\Delta f/f$ in the L-orientation (open circles in Fig.~4). The
difference $-2(\delta f-\Delta f)/f$ yields the additive constant $f_0$.
Given $f_0$ and $\Delta f(T)$ measured in the range $T<T_c$, we also obtain
$\delta f(T)$ in the superconducting state in the L-orientation. As a
result, with due account of
$\lambda_{ab}(T)=\sqrt{1/\omega\mu_0\sigma''_{ab}(T)}$ (insets to Fig.~3),
we derive from Eqs.~(4) and (3) $\lambda_c(0)$, which equals approximately
$\lambda_c(0)\approx 50$~$\mu$m in sample~\#1 and 150~$\mu$m in sample~\#2.
These results are in reasonable agreement with our ac-susceptibility
measurements if we take into consideration the fact that the accuracy of
$\lambda_c$ measurements is rather poor and the error can be up to 30\%.
The temperature dependences of $\sigma''_c(T)$ for both BSCCO single
crystals are shown in Fig.~5.

\begin{figure}[ht]
\centerline{\psfig{figure=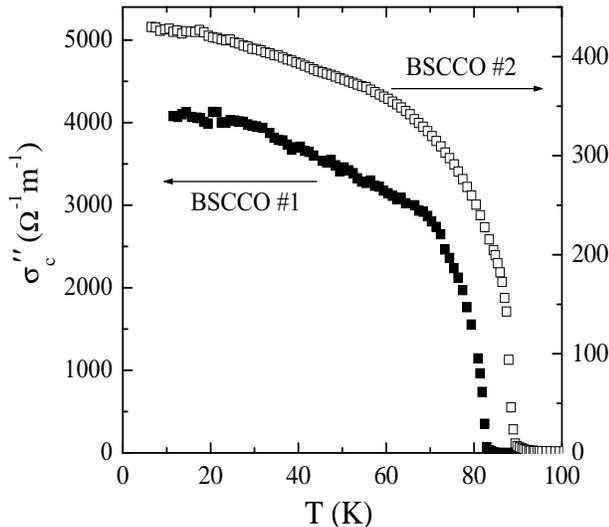,height=7cm,width=8cm,clip=,angle=0.}}
\caption{Conductivities $\sigma''_c(T)$ of BSCCO
single crystals \#1 (left scale) and \#2 (right scale) at 9.4~GHz,
obtained by comparing the measurements of $\Delta(1/Q)$ and
$\Delta f=\delta f-f_0$ to numerical calculations by Equation~(1).}
\label{5}
\end{figure}

\section{microwave absorption measurements in a static magnetic field:
surface barrier and influence of defects}

An alternative technique for the determination of $\lambda_c(0)$ is
based on the measurements of the magnetic field $H_J(T)$
at which Josephson vortices penetrate into the sample.

In the experiment the microwave absorption measured in the L-orientation
as a function of the static magnetic field (0-30~Oe) parallel to the
CuO$_2$ layers exhibits a notable increase at the field $H_J(T)$ \cite{Nic}.
Figure 6 displays the change of microwave dissipation,
starting from zero field
\begin{figure}[h]
\centerline{\psfig{figure=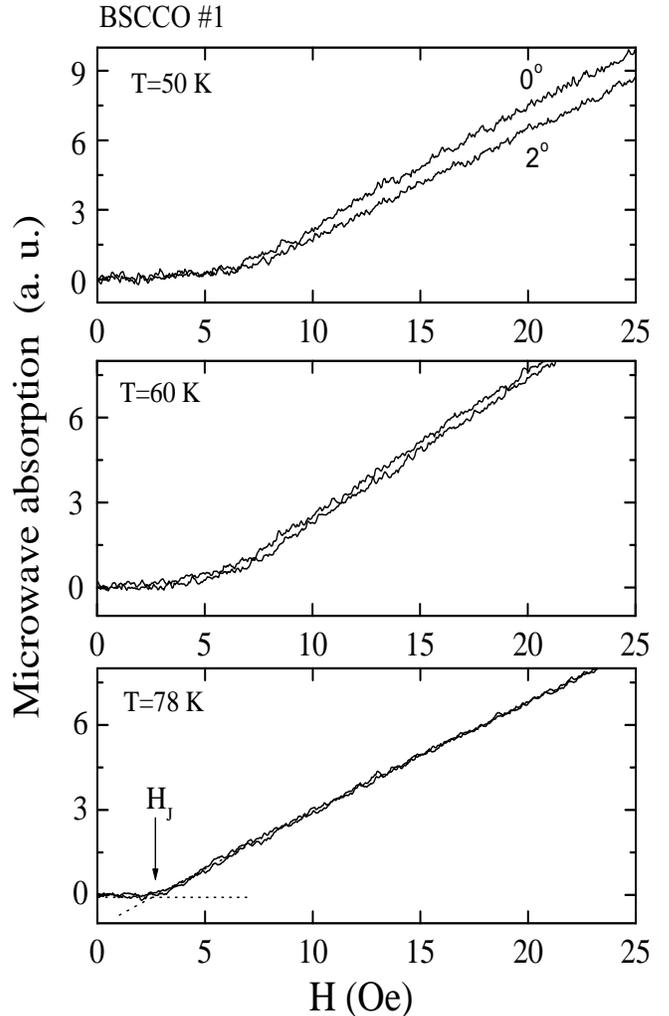,height=13.5cm,width=8.5cm,clip=,angle=0.}}
\caption{Microwave absorption as a function of the applied field $\bf H$
at 3 temperatures, for 2 orientations ($0^{\circ}$ and $ 2^{\circ}$) of the
$\bf H$ with respect to the $ab$-plane. The onset of dissipation, indicated
by the arrow, occurs at the penetration field $H_{J}(T)$.
\label{6}}
\end{figure}
\noindent
(within $\pm 0.1~ \rm Oe$) measured in sample \#1
for various orientations of the applied field close to the $ab$-plane
($0^{\circ}$ and $ 2^{\circ}$) and in a low-field range:
$0~ \leq H \leq 25~\rm Oe$, at three typical temperatures
($T=\rm 78~K, 65~K, 50~K$).
After each field sweep, the sample was warmed through
$T_c$ and then cooled again in zero field, in order to avoid any
possible vortex pinning when studying the penetration starting from zero
field. The dissipation of Josephson vortices is characterized by the fact
that it does not depend on the angle (Fig.~6), as long as these vortices
remain locked. As the field increases, an onset in the dissipation occurs
at a temperature-dependent field $H_{J}(T)$, which we associate to
Josephson vortices entering the sample. As in Ref.~\cite{bontemps}, we
choose to define $H_{J}(T)$ as the field value where the microwave
absorption exceeds the experimental accuracy. The field thus determined
is plotted in Fig.~7. The error bars take into account both the noise and
the estimated drift of the signal with time.
\begin{figure}[h]
\centerline{\psfig{figure=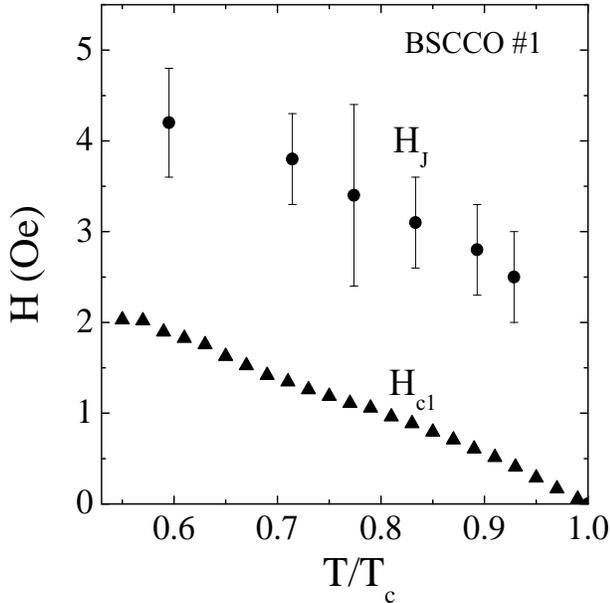,height=8cm,width=8cm,clip=,angle=0.}}
\caption{Plot of $H_{J}(T)$ (full circles) for BSCCO crystal \#1.
Triangles display the upper bond of $H_{c1}(T)$ using
$\lambda_{ab}(0)=2100$~\AA~and $\lambda_{c}(0)=10~\mu$m. The
temperature variations $\Delta \lambda_{ab}(T)$ and $\Delta \lambda_{c}(T)$
are taken from the present work.}
\label{7}
\end{figure}

The magnitude of $H_{J}(T)$
is too large to be associated with the first thermodynamic critical
field $H_{c1}(T)$ for Josephson vortices \cite{clem2}:

\begin{equation}
H_{c1}(T)={{\phi_{0} \over {4 \pi \lambda_{ab}(T) \lambda_{c}(T)}}}
[\ln {\lambda_{ab}(T) /d} + 1.12],
\end{equation}
where $\phi_0$ is the flux quantum and $d$ is the interlayer distance
($\sim 15~\rm \AA$ in BSCCO). Indeed, in Fig.~7 the triangles demonstrate
an upper bound for $H_{c1}(T)$. Here, we take $ 2100$~\AA\ as a lower
bound for $\lambda_{ab}(0)$ \cite{schilling,Lee}, and $10~\mu$m for
$\lambda_{c}(0)$ \cite{tamegai,nakamura,steinmeyer,tsui,khaykovich}.
We use the temperature dependence for $\Delta \lambda_{ab}(T)$
from $\sigma''_{ab}(T)=1/[\omega\mu_0\lambda_{ab}^2(T)]$ shown in the
inset of Fig.~3a,  and $\Delta \lambda_{c}(T)$ (squares in Fig.~2).
It is clearly seen that neither the absolute value
(too small compared to the experimental data) nor the temperature
dependence (quasi-linear) agrees with the $H_{J}(T)$ data.

It is therefore quite natural to assume that a Bean-Livingston
surface barrier \cite{BeLi} impedes magnetic flux penetration into
the sample and yields a higher entry field $H_{\rm SB}(T)$. This assumption
is also supported by the irreversible behavior of the dissipation upon
flux exit \cite{Nic}. In anisotropic superconductors in the quasi-2D regime,
which holds in BSCCO up to temperatures very close to $T_c$, the
field $H_{\rm SB}(T)$ was shown to be related only to the c-axis
penetration length through \cite{buzdin1} :

\begin {equation}
H_{\rm SB}(T)={\phi_{0} \over {4 \pi \lambda_{c}(T) d}}.
\end {equation}

The surface barrier might thus account for the observed value of the
penetration field. Also, since $H_{\rm SB}(T)$ grows as
$1/\lambda_{c}$(T) (instead of $1/\lambda_{ab} \lambda_{c}(T)$),
it is expected that the temperature dependence could show a better agreement.
So we derive from the $H_{J}(T)$ data an effective penetration depth
$\lambda_J(T)=\phi_0/[4\pi H_J(T)d]$ on the analogy of Eq.~(6). The data
are shown in Fig.~8 (full squares). We then try to determine
$\lambda_{c}(0)=\lambda_J(T)-\Delta\lambda_c(T)$ so as to fit
$\lambda_{J}(T)$ using the measured by ac-susceptibility and cavity
perturbation technique
$\Delta \lambda_{c}(T)$ (full circles in Fig.~8 extend the lower
curve in Fig.~2 to higher temperatures). We find that both sets of data,
namely $\lambda_{J}(T)$ and $\Delta \lambda_{c}(T)$, cannot be reconciled
for any value we may assume for $\lambda_{c}(0)$ in the entire temperature
range. Therefore, the interpretation cannot be so simple.
\begin{figure}[h]
\centerline{\psfig{figure=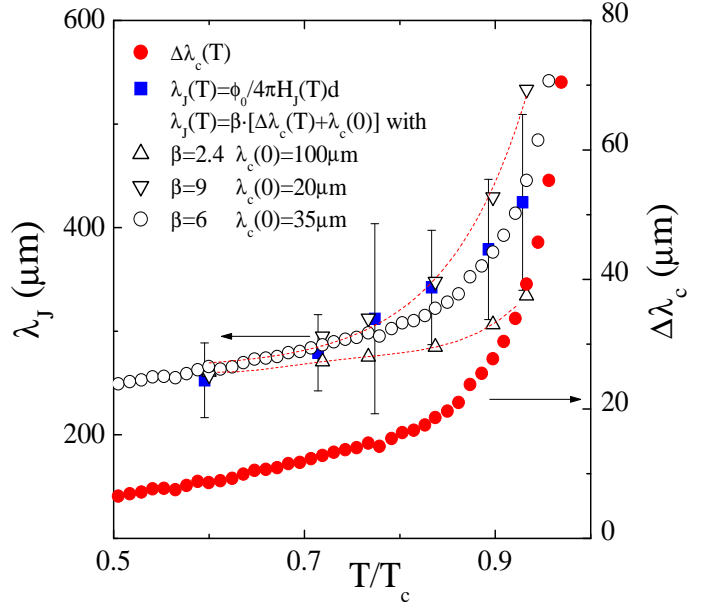,height=8cm,width=9cm,clip=,angle=0.}}
\caption{Plot of the temperature variation $\Delta \lambda_{c}(T)$
(full circles, right scale) and of the effective length $\lambda_J(T)$
which is associated to a surface barrier for the penetration of Josephson
vortices (full squares, left scale) in sample \#1. Open circles
display the best fit using $\lambda_{c}(0)=35~\mu$m and a scaling factor
$\beta=6$. Open symbols show the best fits using $\lambda_c(0)=20~\mu$m
(down triangles) and $\lambda_c(0)=100~\mu$m (up triangles).}
\label{8}
\end{figure}

More detailed quantitative investigation and calculations of the
penetration field in the presence of edge or surface defects, have led
us to the conclusion that $H_{J}(T)$ is eventually controlled by such
surface irregularities \cite{Nic}.

We have carried out several checks (cleaving sample, cutting deep and
wide grooves at the $ab$-plane, rotation crystal along it's $c$-axis)
and established
that large defects, greater or comparable with the penetration depth,
within the $ab$-planes or at the edges of the crystal do not destroy the
surface barrier (no significant change in the onset field of the microwave
absorption was observed). However, if the dimensions of the surface
irregularities are smaller than $\lambda$ then their presence can
strongly influence the screening current distribution. Indeed,
the entrance field is deduced from the balance between the vortex
attraction to the surface and the pushing force exerted by the screening
current at the minimum distance $\xi $ (the vortex core size)
\cite{BeLi,DeGen}. Near a small scratch the current density can
be many times larger than near the flat surface. This may substantially
increase the force pushing vortices inside the superconductor and then
decrease the surface barrier and the entrance field.
As shown in Ref.~\cite{Nic}, in the specific case of a thin groove
with the depth $b\agt d$ at the surface, the penetration field
$H_J(T)=H_{\rm SB}(T)/\beta$, where parameter $\beta=(b/d)^{1/2}>1$.
Based on this estimation, we have attempted to fit our data on
\begin{equation}
\lambda_J(T)={\phi_0\beta \over 4\pi H_{\rm SB}(T)d}=
\beta\left[\Delta\lambda_c(T)+\lambda_c(0)\right],
\end{equation}
using two adjustable parameters $\beta$ and $\lambda_c(0)$ (see Fig.~8).
We have obtained the best fit at $\beta =6$ (or $b\sim 500$~\AA) and
$\lambda_{c}(0)=(35 \pm 15)$~$\mu$m in Eq.~(7), which is in agreement with
our measurements by the first two methods. We also show in Fig.~8 smaller
and larger values for $\lambda_{c}(0)$. They allow us to set the
uncertainty about our determination of the penetration depth.
We can also fit the data in the presence of a slit at the edge of the
crystal \cite{Nic}. The depth of the edge slit should be of the order
of $10~\mu$m, which is still small with respect to $\lambda_{c}(0)$.
The key result in this latter case is that it yields the same absolute
value for $\lambda_{c}(0)$.

\section{conclusion}

In conclusion, we have used three high-frequency techniques
in studying anisotropic properties of BSCCO single crystals.
The results obtained by different techniques are in reasonable agreement.
We have observed almost linear dependences
$\Delta\lambda_c(T)\propto T$ at low temperatures. We have also
determined the absolute value of $\lambda_c(0)$, which is a factor of
three higher in the optimally doped BSCCO sample than in the overdoped
crystal. The ratio between the slopes of curves of $\Delta\lambda_c(T)$
in the range $T\ll T_c$ is the same. These facts could be put down to
dependences of $\lambda_c(0)$ and $\Delta\lambda_c(T)$ on the oxygen
content in these samples. At the same time, the set of our experiments
suggest very strongly the influence of defects in the samples, which
relates the first penetration field  $H_{J}(T)$ of Josephson vortices
to the $c$-axis penetration depth. In order to draw ultimate conclusions
concerning the nature of the transport properties along the $c$-axis in
BSCCO single crystals, studies of more samples with various oxygen
contents are needed.

\section{Acknowledgements}

This work was supported by the Centre National de la Recherche Scientifique
- Russian Academy of Sciences cooperation program 4985, and by CREST
and Grant-in-Aid for Scientific Research from the Ministry of
Education, Science, Sports and Culture (Japan).
The work at ISSP was also supported by the Russian
Foundation for Basic Research (grant 00-02-04021) and
Scientific Council on Superconductivity (project 96060).\\

\end{multicols}


\begin{references}
\bibitem{exp} We note that $\delta f(T)$ includes the frequency shift
due to the sample thermal expansion, which is essential for $T>0.7\,T_c$
in the T-orientation \protect\cite{Tru1}.
\end{references}
\end{document}